# Constant-Weight Gray Codes for Local Rank Modulation

Moshe Schwartz
Electrical and Computer Engineering
Ben-Gurion University
Beer Sheva 84105, Israel
*schwartz@ee.bgu.ac.il*

*Abstract*—We consider the local rank-modulation scheme in which a sliding window going over a sequence of real-valued variables induces a sequence of permutations. The local rank-modulation, as a generalization of the rank-modulation scheme, has been recently suggested as a way of storing information in flash memory.

We study constant-weight Gray codes for the local rank-modulation scheme in order to simulate conventional multi-level flash cells while retaining the benefits of rank modulation. We provide necessary conditions for the existence of cyclic and cyclic optimal Gray codes. We then specifically study codes of weight 2 and upper bound their efficiency, thus proving that there are no such asymptotically-optimal cyclic codes. In contrast, we study codes of weight 3 and efficiently construct codes which are asymptotically-optimal.

## I. INTRODUCTION

In a recent series of papers [26], [27], [37], [39], the rank-modulation scheme was suggested as a way of storing information in flash-memory devices. Basically, instead of a conventional multi-level flash cell in which the charge level of a single cell is measured and quantized to a symbol from the input alphabet, in the rank-modulation scheme the permutation induced by the relative charge levels of several cells is the stored information. The scheme, first described in [26] in the context of flash memory, works in conjunction with a simple cell-programming operation called "push-to-the-top", which raises the charge level of a single cell above the rest of the cells. It was suggested in [26] that this scheme eliminates the over-programming problem in flash memories, reduces corruption due to retention, and speeds up cell programming.

This is certainly not the first time permutations have been used for modulation purposes. Permutations have been used as codewords as early as the works of Slepian [35] (later extended in [2]), in which permutations were used to digitize vectors from a time-discrete memoryless Gaussian source, and Chadwick and Kurz [9], in which permutations were used in the context of signal detection over channels with non-Gaussian noise (especially impulse noise). Further early studies include works such as [2]–[4], [8], [12], [13]. More recently, permutations were used for communicating over powerlines (for example, see [38]), and for modulation schemes for flash memory [26], [27], [37], [39].

An important application for rank-modulation in the context of flash memory was described in [26]. A set of $n$ cells, over which the rank-modulation scheme is applied, is used to simulate a single conventional multi-level flash cell with $n!$ levels corresponding to the alphabet $\{0, 1, \ldots, n! - 1\}$. The simulated cell supports an operation which raises its value by 1 modulo $n!$. This is the only required operation in many rewriting schemes for flash memories (see [5], [23]–[25], [40]). This operation is realized by a Gray code traversing the $n!$ states where, physically, the transition between two adjacent states in the Gray code is achieved by using a single "push-to-the-top" operation.

Most generally, a gray code is a sequence of distinct elements from an ambient space such that adjacent elements in the sequence are "similar". Ever since their original publication by Gray [21], the use of Gray codes has reached a wide variety of areas, such as storage and retrieval applications [10], processor allocation [11], statistics [14], hashing [18], puzzles [20], ordering documents [28], signal encoding [29], data compression [30], circuit testing [31], and more. For a survey on Gray codes the reader is referred to [33].

A drawback to the rank-modulation scheme is the need for a large number of comparisons when reading the induced permutation from a set of $n$ cell-charge levels. Instead, in a recent work [39] the $n$ cells are partially viewed through a sliding window resulting in a sequence of small permutations. We call this the *local rank-modulation scheme*. The aim of this work is to study Gray codes for the local rank-modulation scheme. The paper is organized as follows: In Section II the exact setting, notation, and definitions are presented. We study, in Section III, necessary conditions for the existence of Gray codes for our setting. In Section IV we give constructions for Gray codes of low weight and study their efficiency. We conclude in Section V with a summary and a set of open problems.

## II. DEFINITIONS AND NOTATION

### A. Local Rank Modulation

Let us consider a sequence of $t$ real-valued variables, $\mathbf{c} = (c_0, c_1, \ldots, c_{t-1})$, $c_i \in \mathbb{R}$, where we further assume $c_i \neq c_j$ for all $i \neq j$. The $t$ variables induce a permutation $f_\mathbf{c} \in S_t$, where $S_t$ denotes the set of all permutations over $[t] = \{1, 2, \ldots, t\}$. The permutation $f_\mathbf{c}$ is uniquely defined by the constraints $c_{f_\mathbf{c}(i)-1} > c_{f_\mathbf{c}(j)-1}$ for all $i < j$, i.e., if we sort $\mathbf{c}$ in descending order, $c_{j_1} > c_{j_2} > \cdots > c_{j_t}$ then $f_\mathbf{c}(i) = j_i + 1$ for all $1 \leqslant i \leqslant t$.

Given a sequence of $n$ variables, $\mathbf{c} = (c_0, c_1, \ldots, c_{n-1})$, we define a window of size $t$ at position $p$ to be

$$\mathbf{c}_{p,t} = (c_p, c_{p+1}, \ldots, c_{p+t-1})$$

where the indices are taken modulo $n$, and also $0 \leqslant p \leqslant n-1$, and $1 \leqslant t \leqslant n$.

We now define the *(s,t,n)-local rank-modulation (LRM) scheme*, which we do by defining the *demodulation* process. Let $s \leqslant t \leqslant n$ be positive integers, with $s|n$. Given a sequence of $n$ distinct real-valued variables, $\mathbf{c} = (c_0, c_1, \ldots, c_{n-1})$, the demodulation maps $\mathbf{c}$ to the sequence of $n/s$ permutations from $S_t$ as follows:

$$\mathbf{f_c} = (f_{\mathbf{c}_{0,t}}, f_{\mathbf{c}_{s,t}}, f_{\mathbf{c}_{2s,t}}, \ldots, f_{\mathbf{c}_{n-s,t}}).$$

Loosely speaking, we scan the $n$ variables using windows of size $t$ positioned at multiples of $s$ and write down the permutations from $S_t$ induced by the *local* views of the sequence.

In the context of flash memory storage devices, we shall consider the $n$ variables, $\mathbf{c} = (c_0, c_1, \ldots, c_{n-1})$, to be the charge-level readings from $n$ flash cells. The demodulated sequence, $\mathbf{f_c}$, will stand for the original information which was stored in the $n$ cells. This approach will serve as the main motivation for this paper, as it was also for [26], [27], [37], [39]. See Figure 1 for an example.

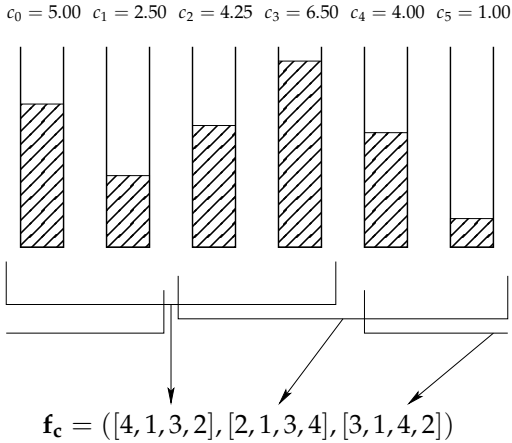

$c_0 = 5.00 \quad c_1 = 2.50 \quad c_2 = 4.25 \quad c_3 = 6.50 \quad c_4 = 4.00 \quad c_5 = 1.00$

$\mathbf{f_c} = ([4,1,3,2], [2,1,3,4], [3,1,4,2])$

**Figure 1.** Demodulating a $(2,4,6)$-locally rank-modulated signal.

We say a sequence $\mathbf{f}$ of $n/s$ permutations over $S_t$ is $(s,t,n)$-*LRM realizable* if there exists $\mathbf{c} \in \mathbb{R}^n$ such that $\mathbf{f} = \mathbf{f_c}$, i.e., it is the demodulated sequence of $\mathbf{c}$ under the $(s,t,n)$-LRM scheme. Except for the degenerate case of $s = t$, not every sequence is realizable.

When $s = t = n$, the $(n,n,n)$-LRM scheme degenerates into a single permutation from $S_n$. This was the case in most of the previous works using permutations for modulation purposes. A slightly more general case, $s = t < n$ was discussed by Ferriera *et al.* [19] in the context of permutation trellis codes, where a binary codeword was translated tuple-wise into a sequence of permutation with no overlap between the tuples. Finally, the most general case was defined by Wang *et al.* [39] (though in a slightly different manner where indices are not taken modulo $n$, i.e., with no wrap-around). In [39], the sequence of permutations was studied under a charge-difference constraint called *bounded rank-modulation*, and mostly with parameters $s = t - 1$, i.e., an overlap of one position between adjacent windows.

Finding out the induced permutation from a sequence of $t$ real-valued readings requires at least $\Omega(t \log t)$ comparisons. Thus, to get the simplest hardware implementation we will consider the case of $t = 2$ throughout the paper. The only non-trivial case to consider is therefore $s = 1$, i.e., a $(1,2,n)$-LRM scheme. Demodulated sequences of permutations in this scheme contain only the permutations $[1,2]$ and $[2,1]$, and a single comparison between the charge levels of two adjacent flash memory cells is required to find the permutation. We will conveniently associate the logical value 1 with the permutation $[1,2]$, and 0 with $[2,1]$, thus forming a simple mapping between length $n$ binary sequences and permutation sequences from the $(1,2,n)$-LRM scheme. It is easily seen that the only two binary sequences not mapped to $(1,2,n)$-LRM sequences are the all-ones and all-zeros sequences.

### B. Gray Codes for $(1,2,n)$-LRM

Generally speaking, a *Gray code*, $G$, is a sequence of distinct states (codewords), $G = g_0, g_1, \ldots, g_{N-1}$, from an ambient state space, $g_i \in S$, such that adjacent states in the sequence differ by a "small" change. What constitutes a "small" change usually depends on the code's application.

Since we are interested in building Gray codes for flash memory devices with the $(1,2,n)$-LRM scheme, our ambient space, which we denote as $S(n)$, is the set of all realizable sequences under $(1,2,n)$-LRM. This is simply the set of all the binary sequences of length $n$, excluding the all-ones and all-zeros sequences, i.e.,

$$S = S(n) = \{0,1\}^n - \{0^n, 1^n\}.$$

Each of the codewords, $g_i \in G$, is a string of $n$ bits which we shall denote as $g_i = g_{i,0}, g_{i,1}, \ldots, g_{i,n-1}$. Throughout the paper we will assume the index $j$ in $g_{i,j}$ is taken modulo $n$, and when appropriate, the index $i$ is taken modulo $N$.

The transition between adjacent states in the Gray code is directly motivated by the flash memory application, and was previously described and used in [26]. This transition is the "push-to-the-top" operation, which takes a single flash cell and raises its charge level above all others.

In our case, however, since we are considering a *local* rank-modulation scheme, the "push-to-the-top" operation merely raises the charge level of the selected cell above those cells which are comparable with it. As the window size is $t = 2$, these cells are the ones directly before and after the selected cell. Thus, we define the set of allowed transitions as $T = \{\tau_0, \tau_1, \ldots, \tau_{n-1}\}$, which is a set of functions, $\tau_j : S \to S$, where $\tau_j$ represents a "push-to-the-top" operation performed on the $j$-th cell. If $v = v_0 v_1 \ldots v_{n-1} \in S(n)$, then $v' = v'_0 v'_1 \ldots v'_{n-1} = \tau_j(v)$ if

$$v'_k = \begin{cases} 0 & k = j \\ 1 & k \equiv j+1 \pmod{n} \\ v_k & \text{otherwise.} \end{cases}$$

Loosely speaking, a transition is made by selecting a window of size 2 in the original codeword, and overwriting it with 01. See Figure 2 for an example.

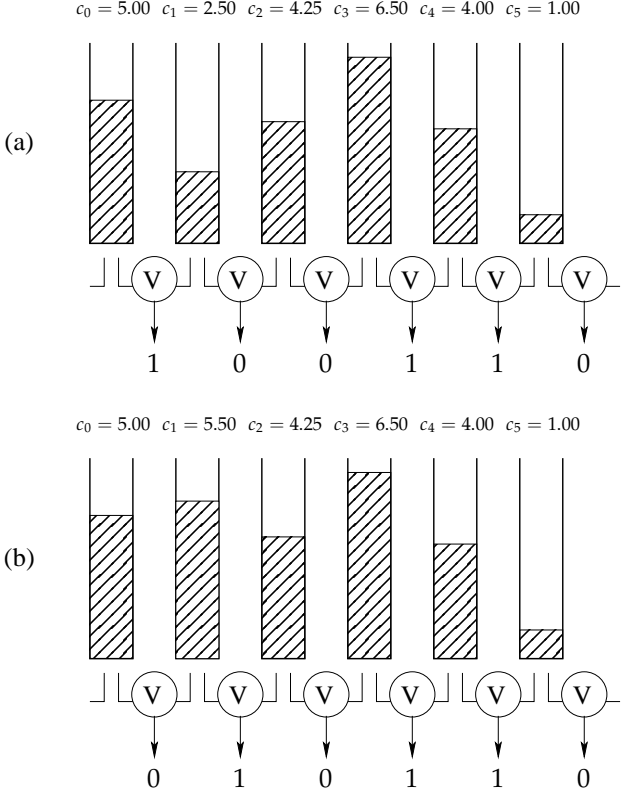

**Figure 2**. An example of a local "push-to-the-top" operation in a $(1, 2, 6)$-LRM scheme. The snapshot (a) presents the system before the change, while (b) presents the system after the change, which locally pushed $c_1$ above $c_0$ and $c_2$, thus changing the first two bits of the demodulated sequence.

**Definition 1.** *A Gray code, $G$, for $(1, 2, n)$-LRM is a sequence of distinct length-$n$ binary codewords, $G = g_0, g_1, \ldots, g_{N-1}$, where $g_i \in S(n)$. For all $0 \leqslant i \leqslant N - 2$, we further require that $g_{i+1} = \tau_j(g_i)$ for some $j$. If $g_0 = \tau_j(g_{N-1})$ for some $j$, then we say the code is cyclic. We call $N$ the size of the code, and say $G$ is optimal if $N = 2^n - 2$.*

When we perform a "push-to-the-top" operation on the $j$-th cell, let us denote its initial charge level as $c_j$, and its resulting charge level as $c'_j$. We set $c'_j$ to $\max\{c_{j-1}, c_{j+1}\} + \delta$, where $\delta > 0$. Two important issues of concern are the difference in charge levels involved in a "push-to-the-top" operation, and cell saturation. In the former, the higher $c'_j - c_j$ is, the more risk of disturbing neighboring cells, while in the latter, the higher we set $c'_j$, the less number of updates to the cell before it saturates. Both concerns benefit from a value of $\delta$ as low as possible. Let us assume that a limited resolution exists and thus $\delta$ is bounded from below by a constant, which w.l.o.g., we can assume is 1 (after a proper scaling).

Let us now assume an optimal setting in which a "push-to-the-top" operation on the $j$-th cell sets $c'_j = \max\{c_{j-1}, c_{j+1}\} + 1$. A general Gray code for $(1, 2, n)$-LRM may result in $c'_j - c_j$ to be exponential in $n$, for some transition from $g_i$ to $g_{i+1}$. The same motivation in the case of $(n, n, n)$-LRM was discussed in [26], where a balanced variant of Gray codes was constructed to avoid the problem. We present a different variant of Gray codes to address the same issue.

First, for any binary string $v = v_0 v_1 \ldots v_{n-1}$, we call the number of 1's in $v$ the *weight* of $v$ and denote it as $\text{wt}(v)$. We also denote by $S(n, w)$ the set of length-$n$ binary strings of weight $w$. We now define our variant of Gray codes:

**Definition 2.** *A constant-weight Gray code for $(1, 2, n)$-LRM, $G = g_0, g_1, \ldots, g_{N-1}$, is a Gray code for $(1, 2, n)$-LRM for which $g_i \in S(n, w)$ for all $i$.*

**Definition 3.** *Let $G$ be constant-weight Gray code for $(1, 2, n)$-LRM with weight $w$ and size $N$. The* efficiency *of $G$ is defined as $\text{Eff}(G) = N / \binom{n}{w}$. If $\text{Eff}(G) = 1$ then we say $G$ is optimal. If $\text{Eff}(G) = 1 - o(1)$, where $o(1)$ denotes a function that tends to 0 as $n \to \infty$, then we say $G$ is asymptotically optimal.*

The transitions between adjacent states in the constant-weight variant take on a very simple form: a window of size 2 in $g_i$ which contains 10 is transformed in $g_{i+1}$ into 01, i.e., "pushing" a 1 one place to the right. Since we are interested in creating cyclic counters, we will be interested in cyclic Gray codes. An example of a cyclic optimal Gray code is given in Table I.

It should be noted that Gray codes with a weaker restriction, allowing a 01 to be changed into 10 and also 10 to be changed back into 01, i.e., a 1 may be pushed either to the right or to the left, have been studied in the past [6], [7], [16], [22], [32].

TABLE I
AN OPTIMAL CYCLIC CONSTANT-WEIGHT GRAY CODE FOR $(1, 2, 5)$-LRM
WITH WEIGHT $w = 2$.

| | | | | |
|---|---|---|---|---|
| 1 | 1 | 0 | 0 | 0 |
| 1 | 0 | 1 | 0 | 0 |
| 0 | 1 | 1 | 0 | 0 |
| 0 | 1 | 0 | 1 | 0 |
| 0 | 0 | 1 | 1 | 0 |
| 0 | 0 | 1 | 0 | 1 |
| 0 | 0 | 0 | 1 | 1 |
| 1 | 0 | 0 | 1 | 0 |
| 1 | 0 | 0 | 0 | 1 |
| 0 | 1 | 0 | 0 | 1 |

We can show that under the constant-weight restriction, for any "push-to-the-top" operation,

$$c'_j - c_j \leqslant \left\lceil \frac{\max\{w, n-w\}}{\min\{w, n-w\}} \right\rceil.$$

This is done by first assuming $2w \leqslant n$, or else we flip all the bits and reverse the codewords. We will only use integer charge levels, and thus for any codeword, $g_i$, $\mathrm{wt}(g_i) = w$, we can find a realization by setting $c_{j+1} - c_j = 1$ if $g_{i,j} = 0$, and $c_{j+1} - c_j = -[(n-w)/w]$ if $g_{i,j} = 1$, where $[\cdot]$ denotes either $\lfloor \cdot \rfloor$ or $\lceil \cdot \rceil$.

It is now easily shown by induction that a "push-to-the-top" operation on the $j$-th cell preserves charge-level differences between adjacent cells and only rearranges their order: by the induction hypothesis, initially we have $c_j - c_{j-1} = -[(n-w)/w]$ and $c_{j+1} - c_j = 1$. The "push-to-the-top" operation sets $c'_j = \max\{c_{j-1}, c_{j+1}\} + 1 = c_{j-1} + 1$ and then $c'_j - c_{j-1} = 1$ and $c_{j+1} - c'_j = -[(n-w)/w]$.

## III. NECESSARY CONDITIONS

We first present a simple necessary condition for the existence of a cyclic Gray code, and then expand it in the case of cyclic optimal codes.

**Theorem 4.** *Let $G$ be a cyclic constant-weight Gray code of size $N$ for $(1,2,n)$-LRM. Then $n|N$.*

*Proof:* We prove the claim using a first-moment coloring argument. For any $v = v_0 v_1 \ldots v_{n-1} \in S(n,w)$, we define the color of $v$ as

$$\chi(v) = \left(\sum_{j=0}^{n-1} j \cdot v_j\right) \bmod n.$$

If $v, v' \in S(n,w)$ and $v' = \tau_j(v)$ for some $j$, then it follows that $\chi(v') \equiv \chi(v) + 1 \pmod{n}$.

Let us now denote $G = g_0, g_1, \ldots, g_{N-1}$. By the previous argument, $i \equiv i' \pmod{n}$ if an only if $\chi(g_i) = \chi(g_{i'})$. Since the code is cyclic, it follows that $N \equiv 0 \pmod{n}$. ∎

We can use Theorem 4 to rule out the existence of cyclic optimal codes in certain cases.

**Theorem 5.** *If $w$ is a prime, then there are no cyclic optimal weight-$w$ Gray codes for $(1,2,n)$-LRM for which $\gcd(n,w) \neq 1$.*

*Proof:* By the assumptions, necessarily $\gcd(n,w) = w$, and so $w|n$. Let $n_w$ denote the exponent of $w$ in the factorization of $n$. We can see that $N = \binom{n}{w} = \frac{n(n-1)\ldots(n-w+1)}{w!}$ and therefore $N_w = n_w - 1$. But then $n \nmid N$ as required by Theorem 4. ∎

The divisibility condition set in Theorem 4 is not strong enough. For example, if we take $n = 12$ and $w = 6$, then indeed $12|\binom{12}{6}$, and the possible existence of a cyclic optimal code with these parameters is not ruled out. However, by the conditions described in the following lemma it is ruled out.

**Theorem 6.** *If a cyclic optimal constant-weight Gray code for $(1,2,n)$-LRM exists, then there are exactly $\binom{n}{w}/n$ strings of each color in $S(n,w)$.*

*Proof:* By Theorem 4 we have $n|\binom{n}{w}$. Furthermore, by the proof of that theorem the code contains an equal number of codewords of each color. Since the code is optimal, i.e., covers all the strings of $S(n,w)$, the claim follows. ∎

To be able to use the last corollary we count the exact number of strings of each color in $S(n,w)$. Though a solution may be deduced from a related theorem due to von Sterneck (see [15], Ch. II), a cleaner self-contained solution follows which is an extension of Sloane's method in [36]. In the following, let

$$A_n(j,k) = |\{v \in S(n,w) \mid \chi(v) = j, k \equiv n - w \pmod{n}\}|$$

for all $0 \leqslant j, k \leqslant n-1$. Also, let $\phi$ stand for Euler's totient function, and $\mu$ stand for Möbius' function.

**Lemma 7.** *The number of strings from $S(n,w)$, $1 \leqslant w \leqslant n-1$, of color $0 \leqslant a \leqslant n-1$, is given by*

$$A_n(a, n-w) = \frac{1}{n} \sum_{\substack{d|n \\ d|w}} (-1)^{\frac{w(d+1)}{d}} \phi(d) \frac{\mu\left(\frac{d}{\gcd(d,a)}\right)}{\phi\left(\frac{d}{\gcd(d,a)}\right)} \binom{n/d}{w/d}.$$

*Proof:* We define the following generating function:

$$f(x,y) = \sum_{j=0}^{n-1} \sum_{k=0}^{n-1} A_n(j,k) x^j y^k.$$

An important observation that follows from the definition of $A_n(j,k)$ is that

$$f(x,y) = \prod_{m=0}^{n-1}(x^m + y) \mod \langle x^n - 1, y^n - 1\rangle.$$

Let $\xi = e^{\frac{2\pi i}{n}} \in \mathbb{C}$ be an $n$-th complex root of unity, then

$$f(\xi^j, \xi^k) = \sum_{j'=0}^{n-1} \sum_{k'=0}^{n-1} A_n(j',k') \xi^{j'j} \xi^{k'k}.$$

Using the inverse two-dimensional discrete Fourier transform we get,

$$A_n(j,k) = \frac{1}{n^2} \sum_{j'=0}^{n-1} \sum_{k'=0}^{n-1} f(\xi^{j'}, \xi^{k'}) \xi^{-j'j} \xi^{-k'k}. \quad (1)$$

Let us denote $g = \gcd(n, j')$. We can directly calculate

$$f(\xi^{j'}, \xi^{k'}) = \prod_{m=0}^{n-1}(\xi^{k'} + \xi^{mj'})$$

$$= (-1)^n \prod_{m=0}^{\frac{n}{g}-1}\left(-\xi^{k'} - \xi^{mj'}\right)^g$$

$$= (-1)^n \left(\left(-\xi^{k'}\right)^{\frac{n}{g}} - 1\right)^g$$

$$= \sum_{m=0}^{g} \binom{g}{m}(-1)^{(g-m)(\frac{n}{g}+1)} \xi^{k'm\frac{n}{g}},$$

where the third equality follows from the well-known fact that $\prod_{i=0}^{n}(z - \xi^i) = z^n - 1$. It now follows that

$$\sum_{k'=0}^{n-1} f(\xi^{j'}, \xi^{k'}) \xi^{-k'k} =$$

$$= \sum_{m=0}^{g} \binom{g}{m}(-1)^{(g-m)(\frac{n}{g}+1)} \sum_{k'=0}^{n-1} \xi^{k'(m\frac{n}{g}-k)}. \quad (2)$$

Since we are interested only in $1 \leqslant k \leqslant n-1$, it follows that $-(n-1) \leqslant k - m\frac{n}{g} \leqslant n-1$, and therefore

$$\sum_{k'=0}^{n-1} \zeta^{k'(m\frac{n}{g}-k)} = \begin{cases} 0 & k \neq m\frac{n}{g} \\ n & k = m\frac{n}{g}. \end{cases}$$

Substituting back into (2), we get for all $1 \leqslant k \leqslant n-1$

$$\sum_{k'=0}^{n-1} f(\zeta^{j'}, \zeta^{k'})\zeta^{-k'k} = \begin{cases} 0 & \frac{n}{g} \nmid k \\ (-1)^{(g-k\frac{g}{n})(\frac{n}{g}+1)}\binom{g}{kg/n}n & \frac{n}{g} \mid k. \end{cases}$$

We again substitute the result back into (1) and summing by divisors of both $n$ and $k$, we get

$$A_n(j,k) = \frac{1}{n^2}\sum_{\substack{d\mid n \\ d\mid k}}(-1)^{\frac{(n-k)(d+1)}{d}}\binom{n/d}{k/d}n\sum_{\substack{m=1 \\ \gcd(m,d)=1}}^{d}\zeta^{-jm}.$$

The inner sum is a Ramanujan sum (see [1]) which equals

$$\sum_{\substack{m=1 \\ \gcd(m,d)=1}}^{d}\zeta^{-jm} = \phi(d)\frac{\mu(\frac{d}{\gcd(d,j)})}{\phi(\frac{d}{\gcd(d,j)})},$$

thus getting

$$A_n(j,k) = \frac{1}{n}\sum_{\substack{d\mid n \\ d\mid k}}(-1)^{\frac{(n-k)(d+1)}{d}}\phi(d)\frac{\mu(\frac{d}{\gcd(d,j)})}{\phi(\frac{d}{\gcd(d,j)})}\binom{n/d}{k/d}.$$

A simple rewriting of the last expression gives the desired result. ∎

The following theorem may be thought of as an extension of Theorem 5 to the case of $w$ not a prime.

**Theorem 8.** *For any fixed weight $w$, there are at most a finite a number of cyclic optimal weight-$w$ Gray codes for $(1,2,n)$-LRM for which $\gcd(n,w) \neq 1$.*

*Proof:* Fix a weight $w$. We will show that there exists $n_0$ such that for all $n > n_0$, $\gcd(n,w) \neq 1$, there is no cyclic optimal Gray code for $(1,2,n)$-LRM. We will do so by showing that $A_n(0, n-w) \neq A_n(1, n-w)$.

Let $p \geqslant 2$ denote the smallest prime number such that $p \mid \gcd(n,w) \neq 1$. We shall also need the fact that

$$\binom{a}{b} = \frac{a}{b}\binom{a-1}{b-1}.$$

Now,

$$A_n(0, n-w) - A_n(1, n-w) =$$
$$= \frac{1}{n}\sum_{\substack{d\mid n \\ d\mid w}}(-1)^{\frac{w(d+1)}{d}}(\phi(d) - \mu(d))\binom{n/d}{w/d}$$
$$= (-1)^{w(p+1)/p} \cdot \frac{p}{n}\binom{n/p}{w/p} +$$
$$\frac{1}{n}\sum_{\substack{d\mid \gcd(n,w) \\ d>p}}(-1)^{\frac{w(d+1)}{d}}(\phi(d) - \mu(d))\binom{n/d}{w/d}.$$

We shall proceed to show that, for large enough $n$,

$$p\binom{n/p}{w/p} > \left|\sum_{\substack{d\mid \gcd(n,w) \\ d>p}}(-1)^{\frac{w(d+1)}{d}}(\phi(d) - \mu(d))\binom{n/d}{w/d}\right|$$

which will prove our claim. Indeed, set $n_0 = \frac{w^3}{2}$, and then for all $n > n_0$

$$\left|\sum_{\substack{d\mid \gcd(n,w) \\ d>p}}(-1)^{\frac{w(d+1)}{d}}(\phi(d) - \mu(d))\binom{n/d}{w/d}\right| \leqslant$$

$$\leqslant \sum_{\substack{d\mid \gcd(n,w) \\ d>p}}(\phi(d) - \mu(d))\binom{n/d}{w/d}$$

$$\leqslant \sum_{d=1}^{w}w\binom{n/p-1}{w/p-1} = \frac{w^3}{n}\binom{n/p}{w/p}$$

$$< 2\binom{n/p}{w/p} \leqslant p\binom{n/p}{w/p},$$

as we claimed. ∎

It should be noted that a more careful analysis can reduce the value of $n_0$ in the proof of Theorem 8. We also observe that when $\gcd(n,w) = 1$, all strings of length $n$ and weight $w$ have full cyclic period. If $v, v' \in S(n,w)$ and $v'$ is a cyclic shift to the right of $v$, then $\chi(v') \equiv \chi(v) + w \pmod{n}$. The fact that $\gcd(n,w) = 1$ also implies that $w$ is a generator of $\mathbb{Z}_n$, and so for every string $v \in S(n,w)$, its $n$ cyclic shifts are all distinctly colored. Thus, $S(n,w)$ has an equal number of strings from each color and the arguments used in the previous theorems will not rule out the existence of cyclic optimal codes.

## IV. Low-Weight Analysis and Constructions

In this section we study constant-weight Gray codes for $(1,2,n)$-LRM having low weight, $w \leqslant 3$ (and by flipping bits and reversing strings, for all $w \geqslant n-3$). In the first trivial case of $w = 1$, there exists a cyclic optimal code for all $n$. As we shall later see, the next two cases $w = 2, 3$ behave radically different: for $w = 2$ we will show that even cyclic *asymptotically-optimal* codes do not exist, while for $w = 3$ we will construct cyclic asymptotically-optimal codes.

### A. The Case of $w = 2$

For the case of $w = 2$ a brute-force approach will suffice. For all $n \geqslant 2$, let us define the graph $\mathcal{G}_n$ whose vertex set is $S(n,2)$ and an edge $v \to v'$ exists iff $v' = \tau_j(v)$ for some $0 \leqslant j \leqslant n-1$.

Since by Theorem 5 cyclic optimal codes may only exist for odd $n$, let us restrict ourselves to that case only. We will, however, specify which results are also valid for even $n$. For convenience, we index the vertices in the following way: $v_{k,\ell}$, where $1 \leqslant k \leqslant (n-1)/2$ and $0 \leqslant \ell \leqslant n-1$, denotes the vertex corresponding to the string having 1's in positions $\ell$ and $\ell + k$. We shall conveniently refer to the first index as the *row* index, and the second index as the *column* index.

Using this indexing method the graph $\mathcal{G}_n$ takes on a simple form for odd $n \geq 5$ (the case $n = 3$ is more degenerate):

- A vertex of the form $v_{1,\ell}$ has a single outgoing edge to $v_{2,\ell}$.
- A vertex of the form $v_{k,\ell}$, $1 < k < (n-1)/2$, has two outgoing edges to $v_{k+1,\ell}$ and $v_{k-1,\ell+1}$.
- A vertex of the form $v_{(n-1)/2,\ell}$ has two outgoing edges to $v_{(n-3)/2,\ell+1}$ and $v_{(n-1)/2,\ell+(n+1)/2}$.

It is now evident that there is a one-to-one correspondence between simple paths in $\mathcal{G}_n$ and Gray codes. A simple construction for an optimal code which is (in general) *not cyclic* is the following.

**Construction 1.** *Let $n \geq 3$ be an odd integer. We construct the following code $G = g_0, g_1, \ldots, g_{N-1}$. We first set $g_0 = v_{1,0}$, and then set $g_{i+1}$ as a function of $g_i = v_{k,\ell}$ according to the following rules:*

- *If $k$ is odd and $k < (n-1)/2$, then $g_{i+1} = v_{k+1,\ell}$.*
- *If $k$ is odd and $k = (n-1)/2$, then $g_{i+1} = v_{k,\ell+(n+1)/2}$.*
- *If $k$ is even and $\ell < n - k/2$, then $g_{i+1} = v_{k-1,\ell+1}$.*
- *If $k$ is even and $\ell = n - k/2$, then $g_{i+1} = v_{k+1,\ell}$.*

**Theorem 9.** *The code from Construction 1 is an optimal constant-weight Gray code for $(1,2,n)$-LRM with $w = 2$.*

*Proof:* It is readily verifiable that the transitions involved in the construction are all valid. Furthermore, the construction is easily seen to first exhaust rows $2t - 1$ and $2t$, where $t \geq 1$, by alternating between them, and then moving to rows $2t + 1$ and $2t + 2$. If the number of rows is even, this is enough to cover all the vertices. If the number of rows is odd, then the last row is covered by transitioning along the row. Since $\gcd((n+1)/2, n) = 1$, $(n+1)/2$ is a generator of $\mathbb{Z}_n$ and the transitions along row $(n-1)/2$ cover all of it. ∎

An example of Construction 1 is shown in Figure 3. When $n = 3, 5$, Construction 1 results in a cyclic code (the case $n = 5$ was given in Table I).

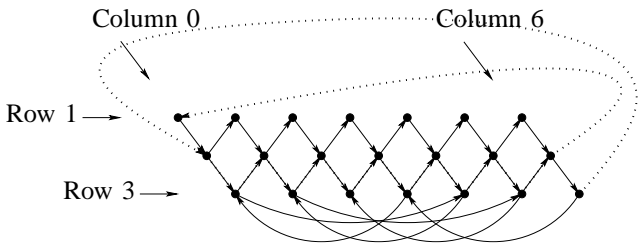

**Figure 3**. An example of an optimal non-cyclic constant-weight Gray code for $(1,2,7)$-LRM with weight 2, which results from Construction 1. Solid arrows represent edges which are part of the code path, while dotted arrows represent those that are not.

**Theorem 10.** *Let $G$ be a cyclic constant-weight Gray code for $(1,2,n)$-LRM with $w = 2$, $n \geq 7$. Then $\mathrm{Eff}(G) \leq \frac{3}{4} + o(1)$.*

*Proof:* We will prove the claim for odd $n$. The proof for even $n$ is essentially the same with a slight difference due to the different structure of $\mathcal{G}_n$. Let $G = g_0, g_1, \ldots, g_{N-1}$ be cyclic Gray code, and let $v_{k_i,\ell_i}$ be the vertex corresponding to $g_i$. We say a vertex $v \in \mathcal{G}_n$ is *covered* if $v = v_{k_i,\ell_i}$ for some $0 \leq i \leq N - 1$. We now denote by $k_{\min}$ and $k_{\max}$ the smallest and, respectively, largest, row index of vertices covered by the code $G$.

The code obviously induces a cyclic path in $\mathcal{G}_n$, and therefore, there exist two sub-paths going "up" and "down" rows, $g_u, g_{u+1}, \ldots, g_{u'}$ and $g_d, g_{d+1}, \ldots, g_{d'}$, with the following properties: (indices are taken modulo $N$ where appropriate)

- $k_u = k_{\min}$, $k_{u'} = k_{\max}$, and for all $0 \leq i \leq (u' - u) \bmod N$, $k_{\min} < k_{u+i} < k_{\max}$.
- $k_d = k_{\max}$, $k_{d'} = k_{\min}$, and for all $0 \leq i \leq (d' - d) \bmod N$, $k_{\min} < k_{d+i} < k_{\max}$.

The two sub-paths are obviously vertex disjoint, except perhaps the first and last vertices of the paths. Furthermore, one can easily be convinced, that the two paths do not occupy the same columns, except perhaps the columns of the first and last vertices of the paths. Along the "up" path, let $0 \leq t_{k_{\min}+1}, \ldots, t_{k_{\max}-1} \leq (u' - u) \bmod N$ be the unique integers such that $g_{u+t_i}$ is the last vertex along the path at row $i$, i.e., $k_{u+t_i} = i$ and for all $t_i < j \leq (u' - u) \bmod N$, $k_{u+j} > i$. It now follows that for each $k_{\min} < i < k_{\max}$, the vertices

$$\left\{ v_{k_{u+t_i}-1,\ell_{u+t_i}+1}, v_{k_{u+t_i}-2,\ell_{u+t_i}+2}, \ldots, v_{k_{\min},\ell_{u+t_i}+k_{u+t_i}-k_{\min}} \right\}$$

cannot be covered by any of the codewords of $G$. See an illustration in Figure 4. The number of such uncovered vertices is exactly $(k_{\max} - k_{\min})(k_{\max} - k_{\min} - 1)/2$.

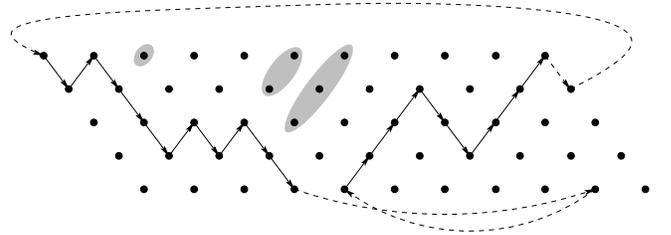

**Figure 4**. An example of a cyclic constant-weight Gray code for $(1,2,11)$-LRM with weight 2. Solid arrows represent edges which are part of the up and down paths, and the shaded vertices are those which are guaranteed to remain uncovered in the proof of Theorem 10.

In addition to the above-mentioned uncovered vertices, all the vertices of rows below $k_{\min}$ and above $k_{\max}$ are left uncovered by definition. Thus, if we denote $\delta = k_{\max} - k_{\min}$, the total number of uncovered vertices is at least

$$n\left(\frac{n-1}{2} - \delta - 1\right) + \frac{\delta(\delta - 1)}{2} \geq \frac{1}{8}(n-3)(n-5),$$

since the minimum is achieved at $\delta = \frac{n-3}{2}$. Therefore, the efficiency of the code $G$ is at most

$$1 - \frac{\frac{1}{8}(n-3)(n-5)}{\binom{n}{2}} = \frac{3}{4} + o(1),$$

as claimed. ∎

While the upper bound on the efficiency presented in Theorem 10 is $\frac{3}{4} + o(1)$, we conjecture that it actually is $o(1)$.

## B. The Case of $w = 3$

In this section we turn to constructing asymptotically-optimal cyclic constant-weight Gray codes for $(1, 2, n)$-LRM with $w = 3$. The construction will use a method originally used for constructing single-track Gray codes in [17] and later in [34]. In fact, the resulting codes will have the single-track property as well.

If $v = v_0 v_1 \ldots v_{n-1}$ is a length $n$ word over some alphabet, let $E$ denote the *cyclic-shift operator* defined by its action on $v$:

$$Ev = v_{n-1} v_0 v_1 \ldots v_{n-2}.$$

The orbits under $E$ are called *necklaces*. A necklace is said to be *full period* if the smallest positive integer $i$ such that $E^i v = v$ is $i = n$. A full-period necklace contains $n$ distinct strings.

We say a Gray code $G = g_0, g_1, \ldots, g_{N-1}$ has the *single-track* property if in the matrix whose $i$-th row is $g_i$, all the columns are cyclic shifts of each other. A variant of the following method was suggested in [17] for constructing single-track Gray codes, and it applies equally well to our set of allowed transitions.

**Lemma 11.** Let $G' = g'_0, g'_1, \ldots, g'_{N'-1}$ be a Gray code for $(1, 2, n)$-LRM where $g'_{i+1} = \tau_{j_i}(g'_i)$ for all $0 \leq i \leq N' - 2$. If the strings in $G'$ are representatives of distinct full-period necklaces, and $E^\ell g'_0 = \tau_{j_{N'-1}} g'_{N'-1}$, $\gcd(\ell, n) = 1$, then the following is a cyclic single-track Gray code:

$$G = G', E^\ell G', E^{2\ell} G', \ldots, E^{(n-1)\ell} G',$$

where $E^j G' = E^j g'_0, \ldots, E^j g'_{N'-1}$.

*Proof:* First, $E^j G'$ is certainly also a Gray code. Since the necklaces in $G'$ all have full cyclic period and since $\ell$ generates $\mathbb{Z}_n$, for $k \not\equiv k' \pmod{n}$ the codes $E^{k\ell} G'$ and $E^{k'\ell} G'$ are disjoint. Finally, it is easy to see that the transition from the last string of $E^{k\ell} G'$ to the first string of $E^{(k+1)\ell} G'$ is valid. ∎

We define the mapping $\psi : S(n, 3) \to \mathbb{Z}_n^3$ as follows: for a binary string $v$ of length $n$ and weight 3 with 1's in positions $0 \leq i_0 < i_1 < i_2 \leq n - 1$, let

$$\psi(v) = (i_1 - i_0, i_2 - i_1, i_0 - i_2)$$

where subtraction is made modulo $n$. The set $\{\psi(v) \mid v \in S(n, 3)\}$ is the set of points $(d_0, d_1, d_2) \in \mathbb{Z}^3$ that are on the hyperplane $d_0 + d_1 + d_2 = n$ restricted to $1 \leq d_0, d_1, d_2 \leq n - 2$. We call $\psi(v)$ the *configuration* of $v$. We note that if $\gcd(n, 3) = 1$, then $S(n, 3)$ contains only full-period strings, and otherwise, all strings are full-period except those with configuration $(n/3, n/3, n/3)$. We denote by $S^*(n, 3)$ the set of full-period strings from $S(n, 3)$.

Since $\psi(v)$, $E\psi(v)$, and $E^2 \psi(v)$, (corresponding to a cyclic rotation of the axes of $\mathbb{Z}^3$), represent strings from the same necklace, for any $v \in S^*(n, 3)$, let $\psi'(v)$ stand for the unique $(d_0, d_1, d_2) \in \{\psi(v), E\psi(v), E^2\psi(v)\}$ for which $d_1 \leq \lfloor n/3 \rfloor$ and $d_2 > \lfloor n/3 \rfloor$. Thus, there is a simple one-to-one mapping from $\{\psi'(v) \mid v \in S^*(n, 3)\}$ to the set of full-period necklaces. We call $\psi'(v)$ the *canonical configuration* of $v$.

A simple counting reveals that there are a total of $\frac{(n-1)(n-2)}{2}$ configurations, and when $\gcd(n, 3) = 1$ there are $\frac{(n-1)(n-2)}{6} = \frac{1}{n}\binom{n}{3}$ canonical configurations which is exactly the number of weight-3 full-period necklaces. When $\gcd(n, 3) \neq 1$, there are $\frac{(n-1)(n-2)-2}{6}$ canonical configurations. See Figure 5 for an illustration.

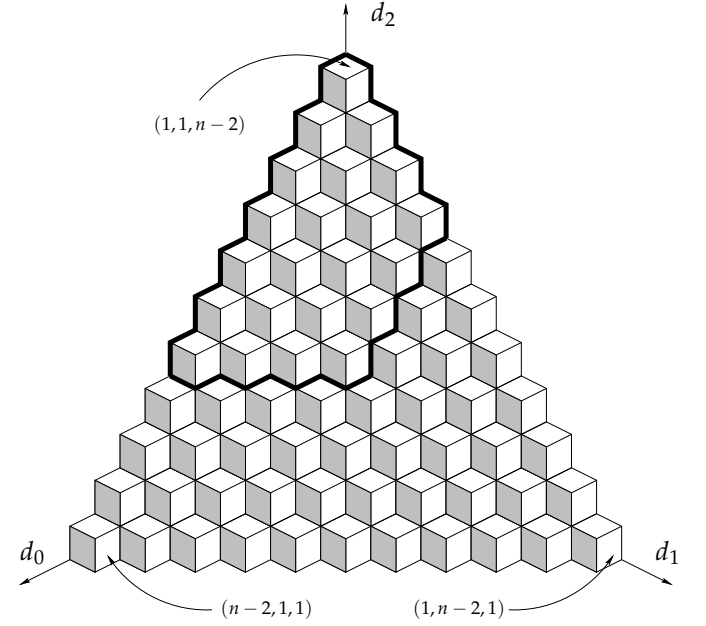

**Figure 5.** The hyperplane of configurations for $n = 13$. The set of canonical configurations is shown surrounded by a thick frame.

**Lemma 12.** Let $\Delta = (d_0, d_1, d_2)$ be a canonical configuration, and assume

$$\Delta' \in \{(d_0 + 1, d_1 - 1, d_2),$$
$$(d_0, d_1 + 1, d_2 - 1),$$
$$(d_0 - 1, d_1, d_2 + 1)\}$$

is also a canonical configuration. Then for any $v \in S^*(n, 3)$ such that $\psi'(v) = \Delta$ there exists $v' \in S^*(n, 3)$ such that $\psi'(v') = \Delta'$ and $v' = \tau_j(v)$ for some $0 \leq j \leq n - 1$.

*Proof:* Assume $\Delta' = (d_0 + 1, d_1 - 1, d_2)$ is a canonical configuration (the proof for the two other cases is similar). Let $v \in S^*(n, 3)$ be such that $\psi'(v) = \Delta$, i.e., there exists some $0 \leq i \leq n - 1$ such that the 1's in $v$ occur in positions $i$, $i + d_0$, and $i + d_0 + d_1$ (all taken modulo $n$). It is easily verified that $v' = \tau_{i+d_0}(v)$ has canonical configuration $\Delta'$. ∎

We now intend to find a long cycle over canonical configurations which, by Lemma 12, will result in a Gray code of representatives of distinct full-period necklaces. The latter will be used with Lemma 11 to generate a cyclic constant-weight Gray code for $(1, 2, n)$-LRM.

**Construction 2.** Let $n \geq 9$ be an integer. We construct the following sequence of canonical configurations $\Gamma = \Delta_0, \Delta_1, \ldots, \Delta_{N'-1}$. We first set $\Delta_0 = (1, 1, n - 2)$, and then

set $\Delta_{i+1}$ as a function of $\Delta_i = (d_0, d_1, d_2)$ according to the following rules:

- If $d_0 = 1$ and $d_1 < 3\lfloor \lfloor n/3 \rfloor /3 \rfloor$, then set $\Delta_{i+1} = (d_0, d_1+1, d_2-1)$.
- Else, if $d_1 \equiv 0 \pmod{3}$, then set $\Delta_{i+1} = (d_0+1, d_1-1, d_2)$.
- Else, if $d_1 \equiv 2 \pmod{3}$ and $d_2 > \lfloor n/3 \rfloor + 1$, then set $\Delta_{i+1} = (d_0, d_1+1, d_2-1)$.
- Else, if $d_1 \equiv 2 \pmod{3}$ and $d_2 = \lfloor n/3 \rfloor + 1$ and $d_1 > 1$, then set $\Delta_{i+1} = (d_0+1, d_1-1, d_2)$.
- Else, if $d_1 \equiv 1 \pmod{3}$ and $d_0 > 2$, then set $\Delta_{i+1} = (d_0-1, d_1, d_2+1)$.
- To complete the cycle, if $\Delta_i = (1, 2, n-3)$, then set $\Delta_{i+1} = (1, 1, n-2)$.

An illustration of the path from Construction 2 is shown in Figure 6.

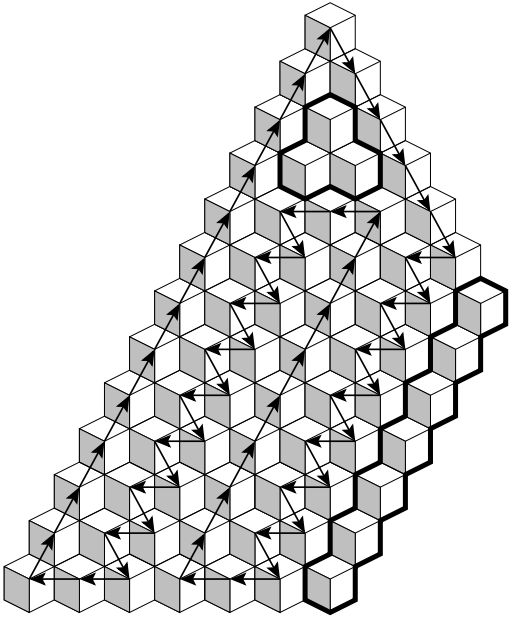

**Figure 6.** The path from Construction 2 over the canonical configurations for $n = 22$. The unvisited configurations are shown surrounded by a thick frame.

**Lemma 13.** *The path from Construction 2 visits only canonical configurations, each visited no more than once.*

*Proof:* Going over all the transitions in Construction 2 one can verify that they visit only canonical configurations. Except for configurations of the form $(1, d_1, d_2)$ which are part of path of increasing $d_1$, the rest of the path is divided according to $d_1 \bmod 3$: when $d_1 \equiv 2, 0 \pmod{3}$ the path zigzags "downward", and goes back "up" when $d_1 \equiv 1 \pmod{3}$ (see Figure 6). This path structure ensures no vertex is visited more than once in a cycle. ∎

**Lemma 14.** *The length $N'$ of the path from Construction 2 is given by*

$$N'(n) = \begin{cases} \frac{n^2-5n+18}{6} & n \equiv 0 \pmod{9} \\ \frac{n^2-5n+22}{6} & n \equiv 1 \pmod{9} \\ \frac{n^2-5n+24}{6} & n \equiv 2 \pmod{9} \\ \frac{n^2-7n+30}{6} & n \equiv 3 \pmod{9} \\ \frac{n^2-7n+30}{6} & n \equiv 4 \pmod{9} \\ \frac{n^2-7n+28}{6} & n \equiv 5 \pmod{9} \\ \frac{n^2-9n+36}{6} & n \equiv 6 \pmod{9} \\ \frac{n^2-9n+32}{6} & n \equiv 7 \pmod{9} \\ \frac{n^2-9n+26}{6} & n \equiv 8 \pmod{9} \end{cases} \quad (3)$$

*Proof:* The path length depends on the number of times it zigzags "downward" which is $\lfloor \lfloor n/3 \rfloor /3 \rfloor$. The rest is a careful and tedious counting. ∎

**Lemma 15.** *Let $G' = g'_0, g'_1, \ldots, g'_{N'-1}$ be a list of strings from $S^*(n, 3)$ (whose existence is guaranteed by Lemma 12) such that $\Gamma = \psi'(g'_0), \psi'(g'_1), \ldots, \psi'(g'_{N'-1})$ is the cyclic path from Construction 2. Let $g^*$ be the string (whose existence is guaranteed by Lemma 12) such that $\psi'(g^*) = \psi'(g'_0)$ and $g^* = \tau_j(g'_{N'-1})$. Then $g^* = E^{N'/3} g'_0$.*

*Proof:* Let us examine $g'_i$ for some $i$ and suppose we could distinguish between the three 1's in $g'_i$ by coloring them red, blue, and green. If $\psi'(g'_i) = (d_0, d_1, d_2)$, assume w.l.o.g., that $d_0$ is the distance between the red and blue 1's, $d_1$ between the blue and green 1's, and $d_2$ between the green and red 1's. If $\psi'(g'_{i+1}) = (d'_0, d'_1, d'_2)$, then a careful reading of Lemma 12 shows that in $g'_{i+1}$, $d'_0$ is again the distance between the red and blue 1's, $d'_1$ between the blue and green 1's, and $d'_2$ between the green and red 1's.

Since $\psi'(g^*) = \psi'(g'_0)$ it follows that $g^*$ is a cyclic shift of $g'_0$. By the previous argument, to get from $g'_0$ to $g^*$, all the 1's had to be pushed an equal number of times to the right and so $g^* = E^{N'/3} g'_0$. ∎

The following is the main theorem of this section:

**Theorem 16.** *For all $n \geq 9$ such that $\gcd(n, N'(n)/3) = 1$, where $N'(n)$ is given by (3), there exists a cyclic constant-weight Gray code for $(1, 2, n)$-LRM with $w = 3$ and size $N = n \cdot N'(n)$, which is also single-track.*

*Proof:* By Lemma 12, let $G' = g'_0, g'_1, \ldots, g'_{N'-1}$ be a list of strings from $S^*(n, 3)$ such that $\Gamma = \psi'(g'_0), \psi'(g'_1), \ldots, \psi'(g'_{N'-1})$ is the cyclic path from Construction 2. By Lemma 14, $N' = N'(n)$ from (3). According to Lemma 13, $\Gamma$ contains distinct canonical configurations, and so $G'$ contains representatives of distinct full-period necklaces. Finally, by combining Lemma 15 with the requirement that $\gcd(n, N'(n)/3) = 1$, we can use Lemma 11 to construct the desired code. ∎

**Lemma 17.** *There are infinite values of $n \in \mathbb{N}$ for which $\gcd(n, N'(n)/3) = 1$. More specifically, it suffices that $n$ satisfies one of the following:*

- $n \equiv 7, 11 \pmod{18}$

- $n \equiv 13, 31, 49, 67 \pmod{90}$
- $n \equiv 5, 23, 41, 59, 95, 113 \pmod{126}$
- $n \equiv 1, 19, 37, 73, 91, 109, 127, 145, 163, 181 \pmod{198}$
- $n \equiv 17, 35, 53, 71, 89, 107, 125, 161, 179, 197, 215, 233 \pmod{234}$

*Proof:* We will prove one of the cases and the rest are similar. Assume $n \equiv 4 \pmod 9$. By Lemma 14 we need

$$\gcd\left(n, \frac{n^2 - 7n + 30}{18}\right) = 1.$$

Since $\gcd(a,b)$ divides any integer combination of $a$ and $b$, and since

$$18 \cdot \frac{n^2 - 7n + 30}{18} - (n-7) \cdot n = 30,$$

it follows that

$$\gcd\left(n, \frac{n^2 - 7n + 30}{18}\right) \bigg| 30.$$

Thus, if we could only make sure that $\gcd(n, 30) = 1$ the claim would necessarily follow. Combining $\gcd(n, 30) = 1$ and $n \equiv 4 \pmod 9$, we get that $n \equiv 13, 31, 49, 67 \pmod{90}$ is sufficient to prove the claim. ∎

We note that the conditions described in Lemma 17 are not the only cases in which $\gcd(n, N'(n)/3) = 1$, but are just the ones easy to derive. For instance, when $n = 27$, we have $\gcd(n, N'(n)/3) = \gcd(27, 34) = 1$.

**Corollary 18.** *There exists an infinite family $\{G_i\}$ of constant-weight Gray codes for $(1, 2, n_i)$-LRM with $w = 3$, $n_{i+1} > n_i$, for which $\mathrm{Eff}(G_i) = 1 - o(1)$.*

*Proof:* Simply combine Lemma 17 with the fact that

$$\lim_{n \to \infty} \frac{n \cdot N'(n)}{\binom{n}{3}} = 1.$$

∎

On a final note, the codes from Theorem 16 turn out to be optimal in the cases of $n = 10, 11$ with sizes $N = 120, 165$ respectively.

## V. CONCLUSION

We presented the general framework of $(s, t, n)$-local rank modulation and focused on the specific case of $(1, 2, n)$-LRM which is both the least-hardware-intensive, and the simplest one to translate between binary strings and permutations. We studied constant-weight Gray codes for this scheme, which guarantee a bounded charge difference in any "push-to-the-top" operation. The Gray codes are used to simulate a conventional multi-level flash cell.

Using coloring and counting arguments we derived necessary conditions for the existence of cyclic and cyclic optimal constant-weight Gray codes for $(1, 2, n)$-LRM.

While cyclic optimal Gray codes exist (trivially) for $w = 1$, we showed that for $w = 2$ their efficiency is upper bounded by $\frac{3}{4} + o(1)$. In contrast, for $w = 3$ asymptotically-optimal codes exist with efficiency $1 - o(1)$. The codes we constructed also come with a relatively simple updating algorithm.

Several open questions still remain. For the case of $(1, 2, n)$-LRM, a general construction is missing for weights $w \geq 4$, and specifically for $w = n/2$ (with the most efficient "push-to-the-top" charge difference). We suspect that for $w = 2$ the efficiency of cyclic codes is actually $o(1)$. Of more general interest is the study of codes for general $(s, t, n)$-LRM and their parameters.